\documentclass[runningheads,a4paper]{llncs}

\usepackage{amssymb}
\usepackage{graphicx}

\usepackage{url}
\urldef{\mailsa}\path| mahyuddin@usu.ac.id|
\urldef{\mailsb}\path| samn@ftsm.ukm.my, saidah@ftsm.ukm.my |
\newcommand{\keywords}[1]{\par\addvspace\baselineskip
\noindent\keywordname\enspace\ignorespaces#1}

\begin{document}

\mainmatter  
% Adaptive Properties for Modelling Simple Search Engine
\title{Social Network Extraction: Superficial Method and Information Retrieval
}

\titlerunning{Social Network Extraction: Superficial Method and ...}
\author{Mahyuddin K. M. Nasution,
\thanks{Proceeding of International Conference on Informatics for Development (ICID'11), c2-110 - c2-115 (2011).}
 Shahrul Azman Mohd. Noah, 
\and 
Saidah Saad
\authorrunning{M. K. M. Nasution, S. A. M. Noah, S. Saad}
\institute{Information Technology Department, \\ Fakultas Ilmu Komputer dan Teknologi Informasi (Fasilkom-TI)\\
Universitas Sumatera Utara, Padang Bulan, Medan 20155, Sumatera Utara, Indonesia\\
\mailsa\\
~\\
Knowledge Technology Research Group, Faculty of Information Science \& Technology\\
Universiti Kebangsaan Malaysia, Bangi 43600 UKM Selangor Malaysia\\
\mailsb\\
}}
\toctitle{}
\tocauthor{}
\maketitle

\begin{abstract} 
Social network has become one of the themes of government issues, mainly dealing with the chaos. The use of web is steadily gaining ground in these issues. However, most of the web documents are unstructured and lack of semantic. In this paper we proposed an Information Retrieval driven method for dealing with heterogeneity of features in the web. The proposed solution is to compare some approaches have shown the capacity to extract social relation: strength relations and relations based on online academic database.

\end{abstract}

\section{Introduction}
Social networks is an approach for representing relations between individuals, groups or organization \cite{nasution2010a}. The concept of social networks extraction is designed to map the relationship of entities among them that can be observed, to mark the patterns of ties between social actors, to measure the social capital: the values obtained by the entities individually or in groups, to present a variety of social structures according  to the interests and its implementation, based on different domains or information sources \cite{nasution2010b}. The extracted social networks can further be processed in information systems, i.e., in data mining systems that detect valuable trends in the social networks, in cases when selecting content based on certain semantic classes, or in expert systems that reason with the extracted knowledge in order to support government decision making. For example, in counterterrorism and the terrorism require networks for learning exchange of information and actions \cite{meter2002}. In scientometric, so extraction of academic social networks aims to see the performance of human resources for example in education \cite{mika2006}.

Web has been chosen as main source to gain various digital information about entities and their relations from all over the world. The web has been becoming the largest text database contained information about social networks, but most of data on the web are so unstructured that they can only be understood by humans, the other way the amount of data is so huge that they can only be processed efficiently by machines. The semantic representation of documents currently forms the vision of semantic web \cite{lee2001}, in the same manner as the social network is a nucleus of semantic relation, i.e. possibilities of attributes partnership between actors where domains generate the attributes and the web documents provide the descriptions of entities and their relations.

Extracting social network is more than the text mining process and has been one of Artificial Intelligent (AI) research agendas, involving the task of identifying entities and their relations for trusted social network. As the manual semantic relation annotation of web documents is impractical and unscalable, and so in the semantic web area there are two research streams for extracting social networks from web documents: unsupervised method and supervised method. The unsupervised methods utilize the Cartesian product for clustering the nodes $A$ in the network. The concept of clustering is $\gamma : A\times A \rightarrow R$ such that $\gamma(a,b) \in R$, $a,b \in A$ \cite{mika2005,matsuo2006,matsuo2007}. However, the clustering approach is difficult to generate the labels of relations in a network. The supervised methods employ a function $\lambda$ for classifying $Z$, i.e. $\lambda : Z \rightarrow C$ such that $\lambda(z) = c$, $z \in Z$, and $c \in C$ is a class label, where $C = \{c_1,c_2,\dots,c_{|C|}\}$ is data set as special target attributes, $|C| \ge 2$ is the number of classes, and $Z\cap C = \emptyset$ \cite{mccallum2005,mccallum2007,tang2008a}. The classification approaches however only concern with extraction of network based on predefined labels only, and thus cannot be adapted to the other descriptions of relations. Therefore, our goal is to enhance the superficial approach, an approach belong to unsupervised method, for extracting social network from web by not only relying on the co-occurrence but to consider other information.

\section{Motivation and Related Works}
An approach for extracting social network is by involving the superficial strategy that depends heavily on the co-occurrence. An occurence, a singleton event of $a$ is a for $a \in A$, $A$ is a set of social actors, ${\bf a} \subset \Omega$, whereas a co-occurrence, the doubleton event of $a$ and $b$, i.e., ${\bf a}\cap{\bf b}$, the subset of $\Omega$, and $a, b \in A$. $\Omega$ is a set of web pages indexed by a search engine, and $P : \Omega\rightarrow [0,1]$ to be an uniform mass probability functions. To make the semantic relation between $a$ and $b$ from Web, we can explore singleton and doubleton, following a pioneer work \cite{kautz1997a} by implementing similarity measure with conditions that $|{\bf a}|\ge |{\bf a}\cap{\bf b}|$ and $|{\bf b}|\ge|{\bf a}\cap{\bf b}|$ \cite{kautz1997b}. At the time of doing this experiment, a Yahoo! search for "Shahrul Azman Mohd Noah", returned 1,200 hits, whereas for "Abdullah Mohd Zin" Yahoo! Search Engine returned 3,870 hits. Searching the pages where both "Shahrul Azman Mohd Noah" and "Abdullah Mohd Zin" = 13 hits, and showed that singletons is greater or equal to doubleton, but we have found that the conditions is not met. Moreover, singletons and doubleton always took along bias in relation, due to the limitations of search engine which due to ambiguity of results.

\subsection{Strength Relation and Similarity (SRS)}
Relations differ in strength. Such strength can be operationalized in a number of ways.  With respect to event frequencies of the pairs of actors where they may exchange large or small amounts of social capital: money, goods, papers, or services. They may supply important or trivial information. Such aspects of relationships measure different types of relational strength. Flink system has been developed to extract, aggregate and visualize a social network \cite{mika2005}. In POLYPHONET system has been created the procedures to expand superficial strategy by providing keywords in query submitted to search engine \cite{matsuo2006}. However, these system took along special cases: Flink for building the social network of a Semantic Web community by utilizing the friend-of-a-friend (FOAF) semantic profile \cite{mika2005}, and POLYPHONET developed to identify the relations in the Japan AI conference \cite{matsuo2007}. 

The content of a relation refers to the resource that is exchanged that may generate some labels of relation. In this case, each entity will be assigned with multiple labels extracted from information sources related to that entity. Suitable labels will be assigned to relations by using the Information Retrieval technique, mainly the generative probabilistic model (PGM) \cite{mccallum2004}. The parameters of GPM are used as modalities to get the knowledge from the corpus, and generating strength relation based on labels of entities. For example, the strength relation between entities based on participants in the same conference or workshop \cite{tang2008b}.

Similarity measures the relation between two entities are a strategy to generate strength relation. One of the used similarity measure widely is the Jaccard coefficient, in singleton and doubleton we have
\begin{equation}
sim_{jac}(a,b) = \frac{|{\bf a}\cap{\bf b}|}{(|{\bf a}|+|{\bf b}|-|{\bf a}\cap{\bf b}|)}
\end{equation}

\subsection{Underlying Strength Relation (USR)}
Underlying strength relation is an approach to exploit URL (Uniform Resource Locator) addresses and its organization since URL address is always available in Web snippets, returned by any search engine \cite{nasution2010a}. Syntactically, an URL represents a resource in Internet. A composition of URL contains a set of tokens, $U = \{s, d_1,\dots,d_m, p_1,\dots,$ $p_{n-1}\}$, satisfying a structure: $s://d_m.\dots .d_2.d_1/p_1/p_2/$ $\dots/p_{n-1}$, i.e. a string consists of scheme, authority, and path. The scheme is a token s, a component contains a protocol that is used for communicating in Internet. For example, http (Hypertext Transfer Protocol), https (http Security) and other protocols. The authority is string of tokens as $dm.\dots .d_2.d_1$, i.e. a component has three subcomponents: user information, host, and port:
\begin{enumerate}
\item The user information may consist of a user name and, optionally, scheme-specific information about how to gain authorization to access the resource. Usually, it is followed by a commercial at-sign("@") that delimits it from the host, if present like in an email address \url{mahyuddin@usu.ac.id}. 
\item The host contains a location of a web server, where the location can be describe as either a domain name system (DNS) or internet protocol (IP) address.
\item The port is a specific number. For instance, a default port number (80 for the http protocol), i.e. $s://d_m.\dots.d_2.d_1:80/$. The colon symbol (":") should be prefixed prior to the port number. 
\end{enumerate}

The last string of tokens $d_1/p_1/p_2/\dots/p_{n-1}$ is a path, i.e. a component contains the directories including a web page and a file name of the page, where a directory and a file are separated by the slash symbol ("/"). The last token of path sometimes comes with two other components: query and fragment. For example, \url{http://search.yahoo.com/search;_ylt=AjoEJrO9wuxK84pfA74_RvCbvZx4?vc=&fp_ip=my&p=Mahyuddin+K.+M.+Nasution&toggle=1&cop=mss&ei=UTF-8&fr=yfp-t-701}. The query is a component containing parameter names and values that may to be supplied to web applications. The token of path and the query are separated by the question symbol ("?"). The form of query is name=[value], where there is equal symbol ("=") between a parameter name and a parameter value. A pair of name=[value] is separated each other by the ampersand symbol ("\&"). The fragment is a component for indicating a parameter part of a document. This last component and the part previously mediated by a sharp symbol ("\#"). Consequently, be found same URL addresses, but presented in a different threads. Therefore, necessary to the canonicalization of the URL. 

URL address indicates the layered structure of a web site which can be logically shown as a hierarchy. As such the URL of web pages which provides and indicator of its logical position in the hierarchy structure that can be considered as the underlying strength of the relationship, where site editors usually tend to put similar or related web pages as close as possible underlying relations among entities in the case that co-occurrence measures unable to provide such relations. 

For any web snippet produces as a results of the entity name query, there exists a set of $k$ URL addresses. Therefore, there will be $n_1+\dots+n_k$ URL addresses whereby $n_i$, $i=1,\dots,k$, is the number of layers for each $i$. For these generated URLs, there is a possibility of redundant URLs. Let u is the number of same URL address. For each entity $a \in A$ we can derive a vector space $a = [a_1,\dots,a_K]$ where $a_j = un_i$. We, therefore, can measure the distance between the two entities based on the list of URL addresses from Web snippets \cite{nasution2010a}.

\subsection{Associaton Rule and Similarity (ARS)}
Another implementation of co-occurrence is a formulation of basic data mining, e.g. the association rule \cite{nasution2011}. Assume $B = \{b_1,b_2,\dots,b_{|B|}\}$ is a set of attribute literals, and a set of transaction $M_i$ are subsets of attributes, or $M_i$ be subsets of $B$. Then, we define the implication, $X\Rightarrow Y$ with two possible values $T = \textsc{TRUE}$ or $F = \textsc{FALSE}$ as association rule if $X$ be a subset of $B$, $Y$ be a subset $B$ and $X\cap Y\ne\emptyset$. Let $q$ = "$a$ AND $x$" is a query, where $x$ is a keyword and $a$ is a name of actor $a \in A$ as seeds, and $Db_i$ is a collection of document containing names of actors $b_i \in A$, then the transactions be $M_i = \{q,b_i\}$ or $\{(q \Rightarrow b_i)\}$, $q \in X$, $b_i \in Y$, $M_i \in M$. So, by making q always T, (see Table 1), we obtain a conditional probability as follows.

\begin{table}
\caption{Transaction and implication}
\centering
\begin{tabular}{|c|c|c|c|c|c|}\hline
Transaction & \multicolumn{2}{|c|}{$~~q=(t_a,t_x)~~$} & $~~t_{b_i}~~$ & \multicolumn{2}{|c|}{Implication}\cr\hline
 $M_i$      & $~~a~~$ & $t_x$ & $b_i$ & $a\Rightarrow t_x$ & $q\Rightarrow b_i$\cr\hline
$M_1$ & T & T & T & T & T\cr
$M_2$ & T & T & $\dots$ & T & $\dots$\cr
$\vdots$ & $\vdots$ & $\vdots$ & $\vdots$ & $\vdots$& $\vdots$\cr
$M_{|M|}$ & T & T & $\vdots$ & T & $\vdots$\cr\hline
\end{tabular}
\end{table}
\begin{equation}
p(b_i|a) = \frac{|{(q\Rightarrow b_i)=T}|}{|M|}.
\end{equation} 		

The Jaccard coefficient in Eq. (1) is modified as follows.
\begin{equation}
sim(a,b_i) = \frac{|{(a\Rightarrow b_i)=T}|}{(|M|+|Db_i|-|{(q\Rightarrow b_i)=T}|)}. 	
\end{equation}

We used the association rule for extracting social network from online database such as DBLP and for enhancing the superficial method, but this approach depend on the structure of DBLP web page. In previous research, association rule can be generally defined as $b_1,\dots,b_{|B|-1} \Rightarrow b_{|B|}$, where $X = \{b_1,\dots,b_{|B|-1|}\}$, and $Y = \{a_{|B|}\}$. If we used Table 1 to generate labels of networks in a tree, then we can use TF.IDF (term frequency - inverse document frequency) scheme for extracting label by considering $\sigma =$ degree of node as the tree root, i.e.
\begin{equation}
\begin{array}{rcl}
TF.IDFw &=& tf(w)\cdot idf(w)\cr
        &=& \Big(\sum_{j\in\{1,\dots,N\}}\sum_{i\in\{1,\dots,m\}}\frac{1}{n}\Big) \log\frac{N}{df(w)}\cr
\end{array}
\end{equation}
where $n$ is the number of words in a document, $m$ is the number of word $w$ in document $j$, $N$ is the total number of documents, and $df(w)$ is the number of documents containing the word $w$. The normalization of TF.IDF, is defined as 
\begin{equation}
tfidf_{nor} = (TF.IDF)(N/\sigma)
\end{equation}

\section{Extraction and Information Retrieval}
First step for building any network is to determine nodes \cite{lin1999,newman2003}. Discretely, a node $v \in V$ in a graph $G$, is a representation of any object in a network, i.e. $V = \{v_i|i = 1,\dots,n\}$, $V \ne\emptyset$. The nodes in a social network refer to actor names such as authors, recipients, researchers, artists. Therefore, the first task of extracting social network, $\xi$, is to identify the actors. This can be achieved by providing a list of names as seeds in order to extract other names, recognize and disambiguate them \cite{balog2009,cullota2004}. The actors play some role in a social based on their background, and they have some characteristics as attributes. Formally, such attributes of actors we define as $Z = \{z_j|j=1,\dots,m\}$ which are attributes / characteristics of entities whereby a pair of $\langle A,Z\rangle$ is the instance of actors, where $Z_i$ are subsets of $Z$, $Z_i$ are subsets of attributes of each entity $a_i$, i.e. $\langle a_i,Z_i\rangle$, $i = 1,\dots,n$, or simply denotes a set of attributes of entity a as $Z_a$.

We developed an approach to extracting keywords from web snippets for disambiguating names in social networks. The methods for using keyword in query are \cite{mika2007}:

\begin{tabular}{rcl}
noK	&:& A name pair  used as query without relation keywords.\cr
K1  	&:& A name pair and top-weighted relation keyword used as a query.\cr
K2  	&:& A name pair and second-weighted relation keyword used \cr
        & & as a query.\cr
K1+K2 	&:& A name pair and top-weighted strongest relation keyword.\cr
\end{tabular}

\noindent
By considering a set of snippets $L$, each contains a reference to a person. Let $P = \{P_1,\dots,P_{|P|}\}$ be a partition of $\xi$ and $\zeta$ into references to the same person, so for example $P_i = \{S_1, S_4, S_5, S_9\}$ might be a set of references to "Abdul Razak Hamdan" the information technology professor. We produce the current context for each actor where keywords have two vector, $\nu$ and $\upsilon$: first vector we define based on TF.IDF and second vector we generate from hit count, and then we define delta $\delta$ as distance of two vector. We use delta to select keyword from each classes of candidates that grouped by tree of network semantic of words \cite{nasution2010a}.  

Let $C = \{C_1,\dots,C_{|C|}\}$ be a collection of disjoint subset of $L$ created by algorithm and manually validated such that each $S$ has an identifier, i.e. URL address. Then, we denote $L_C$ as references that  be the clusters based on collection. Based on measure were introduced, we define an notation of recall $Rec()$ as follows 
\begin{equation}
Rec(S_i) = \frac{|\{S \in P(S_i) : C(S) = C(S_i)\}|}{|\{S \in P(S_i)\}|}
\end{equation}
and a notation of precision $Prec()$ as follows
\begin{equation}
Prec(S_i) = \frac{|\{S \in C(S_i) : P(S) = P(S_i)\}|}{|\{S \in C(S_i)\}|}
\end{equation}
where $P(S_i)$ as a set $P_i$ containing reference $S_i$ and $C(S_i)$ to be the set $C_i$ containing $S_i$. Thus, the precision of a reference to "Abdul Razak Hamdan" is the fraction of references in the same cluster that are also to "Abdul Razak Hamdan". We obtain average of: recall (REC), precision (PREC), and F-measure for the clustering C as follows:
\begin{equation}
REC = \frac{\sum_{S\in L_C} Rec(S)}{|L_C|} 
\end{equation}
\begin{equation}
PREC = \frac{\sum_{S\in L_C} Prec(S)}{|L_C|}
\end{equation}
\begin{equation}
F = \frac{2\cdot REC\cdot PREC}{REC+PREC}
\end{equation}

Second step, for building any network is to determine an edge based on a concept of graph theory \cite{newman2003}. In social network, the edges refer to relationships between actors. Therefore, another main objective of extracting social network is to identify relation among entities, i.e., second task of social network extraction, $\zeta$. If there exists $R = \{r_1,\dots,r_k\}$ as a set of possible relations between social actors, then we have $R = A \rightarrow A\times A = r(a,b)$, $\forall a,b \in A$, that can be depicted as the overlap principle, i.e. the intersection their attributes $r_k(a,b) = Z_a\cap Z_b$. This means that the relations among actors are formed by sharing attributes, ideas, and concepts between them. 

Social network extraction provides the technology to identify and describe content. The technology is as an exertion to further acquire rich and trusted social network. Such social network formally is $SN = \langle V,E,A,R,Z,\xi,\zeta\rangle$, $V \ne\emptyset$, $A \ne\emptyset$ that satisfies the following conditions:
\begin{enumerate}
\item $\xi : A\stackrel{1:1}{\rightarrow} V$, $v = \xi(a)$, for every $a \in A$ there is only one $v \in V$;
\item $\zeta : R \rightarrow E$ so that $e_j = \zeta(r_k(a,b)) = \zeta(Z_a\cap Z_b)$, $e_j \in E$, $r_k \in R$, $\forall a,b \in A$, where $Z_a, Z_b, Z_a\cap Z_b\subset Z$.
\end{enumerate}

The labeled social network is to present a variety of social structures according to the interests and its implementation. The strength relation in social network tends to ambiguity. One or more relations may connect a pair of actors, if it applies, we have social network with many ties. Pairs may maintain a tie based on one relation only, e.g., as members of the same organization, or they may maintain a multiplex tie, based on many relations, such as sharing information, giving financial support and attending conference together. Semantically, the composition of a relation or a tie is derived from the social attributes of both participants, for example is the tie between different or same sex dyads, between a supervisor and an underling or between two peers. Computer network tends to underplay the social cues of participants than the connections, but once a computer network is social network. Once again, we define the social network as $SN_{des} = \langle V,E,A,R,Z,\xi,\zeta,\varphi_1,\varphi_2\rangle$, $V\ne\emptyset$, $A\ne\emptyset$, a labelled social network satisfying the following conditions: 
\begin{enumerate}
\item There exists $\varphi_1 : Z \rightarrow V$ so that $\forall z_i \in Z_a$, $\varphi_1(z_i) \in V$, and $\forall a \in A$ and $Z_a\subset Z$; 
\item There exists $\varphi_2 : Z \rightarrow E$ so that $\forall z_j \in Z_a\cap Z_b$, $\varphi_2(z_j) \in E$, $\forall a,b \in A$, and $Z_a, Z_b, Z_a\cap Z_b \subset Z$.
\end{enumerate}

Extracting social network from web documents need technologies. Those technologies will become an important component of any retrieval system or as Information Retrieval (IR) where the representations of documents and query are enriched with semantic information. From social network perspective, it will give emersion that can adapted and advance the retrieval models, i.e. the representations of query and document relations and by the function that estimates the relevance of relations to a query. Therefore, we can consider relationship extraction as an information retrieval task. Let $G_1 = \langle V_1,E_1\rangle$ is a graph as the resulted social networks and $G_2 = \langle V_2,E_2\rangle$ is a benchmark graph,  we have Jaccard-coefficient, or Eqs. (1) and (8)-(10) be \cite{mika2007} 
\[
sim_G = \frac{|E_1\cap E_2|}{|E_1|+|E_2|-|E_1\cap E_2|},
\]
\[
Precision = \frac{|E_1\cap E_2|}{|E_1|},
\]
\[
Recall = \frac{|E_1\cap E_2|}{|E2|}
\]
and 
\[
F_{measure} = \frac{2(|E_1\cap E_2|)^2}{|E_1|\cdot|E_1\cap E_2|+|E_2|\cdot|E_1\cap E_2|}.
\]
Information retrieval is concerned with answering information needs as accurately as possible, where relations retrieval typically involves the querying of structured data from unstructured information.  

\section{Result}

\subsection{Evaluation and Dataset}
For evaluation of the approaches, we have gathered and labeled  a dataset of 539 Web pages, Tabel 2, where the label we created based on URL address of each web page in snippet. 

\begin{table}
\caption{Statistics of our dataset}
\centering
\begin{tabular}{|l|c|r|}\hline
\multicolumn{1}{|c|}{Personal name}|	& Position & \multicolumn{1}{|c|}{Number of pages}\cr\hline
Abdul Razak Hamdan & Dean   & 85\cr
Abdullah Mohd Zin & Professor & 90\cr
Shahrul Azman Mohd Noah & Professor & 134\cr
Tengku Mohd Tengku Sembok & Professor & 189\cr
Md Jan Nordin &	Assoc. Prof.	& 41\cr\hline
\multicolumn{2}{|c|}{Total} &	539\cr\hline
\end{tabular}
\end{table}

As banchmark of social network, we possess $|V_2| = 67$ nodes and $|E_2| = 253$ edges. This graph we derived from online database DBLP to 67 academic persons, we extract by using association rule (ARS) and then evaluated and corrected their relations based on author-coauthor relationship.
\begin{figure}
\centering
\includegraphics[height=6.5cm]{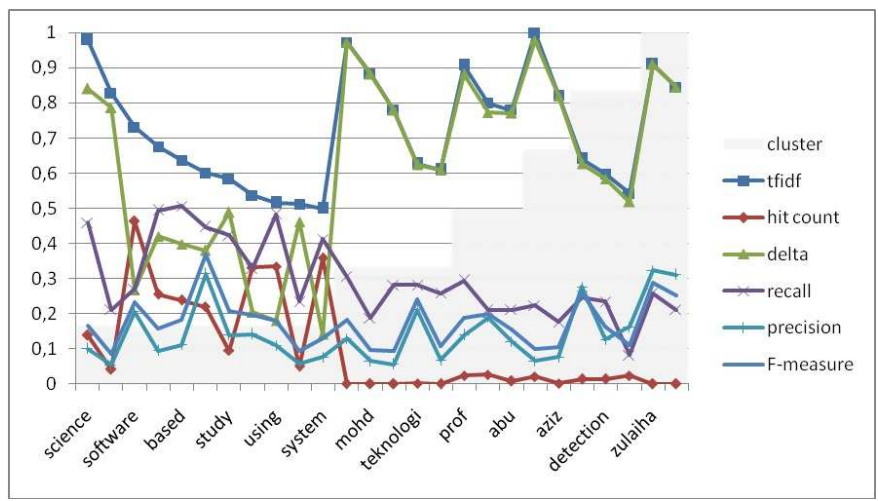}
\caption{(TF.IDF, hit count, delta) vs (recall, precision, F-measure) of candidate words for actor "Abdul Razak Hamdan".}
\label{fig:matrix}
\end{figure}
\subsection{Experiment}
Let us consider information context of actors that includes all relevant relationships as well as interaction history, where Yahoo! Search engine fall short of utilizing any specific information, especially context information, and just use full text index search in web snippets. In experiment, we use maximum of 600 web snippets for search an actor a, and we consider words where the TF.IDF value $> 0.3 \times$ highest value of TF.IDF, or maximum number is 30 words, see Fig. 1. We used Eqs. (6), (7) and (10) for computing recall, precision and F-measure.

\begin{table}
\caption{Name disambiguation result}
\centering
\begin{tabular}{|c|c|c|c}\hline
Method & Recall & Precision & F-measure\cr\hline
Delta ($\delta$) & 45.8 & 29.5 & 35.9\cr\hline
\end{tabular}
\end{table}

Under recall, precision and F-measure: Eqs. (8)-(10), and results in Table 3, this method shows something to consider, i.e. the number of words in the cluster should be limited so that an average value of measurement is not affected by the lower (see Fig. 1). Given the implementation of this method is done with the Yahoo! search, this result is still reasonable.

We tes for 213 actor names and there are 22,683 potential relations, but at the time of doing this experiment by using SRS there are only 12,621 (53\%) relations which satisfy threshold $\alpha = 0.0001$. However, the method (USR) that involves URL-computation able to identify 19,513 (86\%) relations for $\alpha = 0.01$.

\begin{table}
\caption{Recall and precision of social network}
\centering
\begin{tabular}{|c|c|c|c|}\hline
Method & Recall & Precision & F-measure\cr\hline
SRS & 120/253 (47\%) & 120/12,621 (10\%) & 27\%\cr
URS & 176/253 (70\%) & 176/19,513 (9\%) & 18\%\cr\hline
\end{tabular}
\end{table}

Tabel 4 shows the result of correlation with the SRS and USR where based on Jaccard coefficient $sim(SRS,ASR) = 0.0094$ almost the same with $sim(URS,ASR) = 0.009$, but looking at those results USR shows good performance.

\section{Conclusion and Future Work}
The method used URL as object for extracting social network to be incorporated into existing tracting method of social network. It shows how to uncover underlying strength relations by exploiting Web snippets and URL structure. This showed that a well-known paradigm of querying a document web more simple for accessing by inputting keyword, but most simply by using URL address, because on the word or keyword we have typical problems of synonymy and ambiguity, but URL not exist. Our near future work is to further experiment he proposed method and look into the possibility of enhancing IR performance by using social network with developing banchmark manually.

\end{document}